\documentclass[lettersize,conference]{IEEEtran}
\IEEEoverridecommandlockouts

\usepackage{cite}
\usepackage{amsmath, amssymb, amsfonts}
\usepackage{algorithmic}
\usepackage{graphicx}
\usepackage{subcaption}
\usepackage{textcomp}
\usepackage{xcolor}
\usepackage{bm}
\usepackage{bbm}
\usepackage{float}
\usepackage{url}
\usepackage{booktabs}
\usepackage{balance}

\usepackage{tikz}
\usetikzlibrary{positioning}
\usepackage{tkz-graph}
\usepackage{pgfplots}
\usepackage{pgfplotstable}

\usetikzlibrary{math, calc, positioning, shapes, backgrounds, fit, arrows, arrows.meta, decorations.pathreplacing}
\pgfplotsset{compat=1.18}

\def\BibTeX{{\rm B\kern-.05em{\sc i\kern-.025em b}\kern-.08em T\kern-.1667em\lower.7ex\hbox{E}\kern-.125emX}}

\newcommand{\brc}[1]{\left\{#1\right\}}

\newcommand{\hhquad}{\hspace{0.25em}}

\DeclareMathOperator{\sign}{sign}
\DeclareMathOperator*{\argmin}{arg\,min}

\begin{document}

\title{A Log-Domain Approximation of SOCS Decoding for Turbo Product Codes}

\author{
  \IEEEauthorblockN{Oleg Nesterenkov, Kirill Andreev, Alexey Frolov, Pavel Rybin}

  \IEEEauthorblockA{\textit{Center for Next Generation Wireless and IoT} \\
  \textit{Skolkovo Institute of Science and Technology}\\
  Moscow, Russia \\
  \{o.nesterenkov, k.andreev, al.frolov, p.rybin\}@skoltech.ru}
}

\maketitle

\begin{abstract}
This paper studies low-complexity soft-output decoding of turbo product codes with extended Bose–Chaudhuri–Hocquenghem component codes. Recent soft-output from covered space (SOCS) decoding substantially improves the quality of extrinsic information compared with the conventional Chase--Pyndiah decoder, but its probabilistic-domain implementation is less attractive for hardware-oriented realizations. We therefore propose a log-domain approximation of SOCS based on max-log approach. The proposed soft-input soft-output rule replaces probability-domain operations with a piecewise-linear function of reliability gaps between competing Chase-II decoding list and out of the list hypotheses, which preserves compatibility with the standard iterative TPC decoding loop. Numerical results for a TPC built from $(256,239)$ eBCH component codes show that the proposed decoder clearly outperforms the baseline Chase--Pyndiah decoder with the same list size and approaches the performance of SOCS decoder.
\end{abstract}


\section{Introduction}
\label{sec:intro}

Modern communication and storage systems require forward error correction schemes that simultaneously provide high coding gain, low error floors, and hardware-efficient implementation. Turbo product codes (TPCs), originally introduced in~\cite{Elias}, remain attractive in this context because they combine strong performance with a highly regular decoding structure. In particular, TPCs with algebraic component codes are actively used and studied for high-throughput communication and storage scenarios due to the availability of efficient iterative decoders and the possibility of massively parallel row--column processing on FPGA and ASIC platforms~\cite{Mukhtar2016, FPGA, ASIC}.

An important advantage of TPCs with algebraic component codes, such as Bose--Chaudhuri--Hocquenghem (BCH) codes, is that their error-correcting capability can often be characterized analytically or at least bounded with high confidence~\cite{Justesen2011}. In the purely hard-decision setting, this property enables one to predict the error-rate behavior of the decoder without relying exclusively on extremely long Monte Carlo simulations.

At the same time, the reliability and spectral-efficiency targets of modern systems can no longer be met with purely hard-decision decoding. Achieving competitive performance requires iterative soft decoding, which in the TPC setting means that every component decoder must operate in a soft-input soft-output (SISO) mode. Therefore, the central algorithmic problem is to construct a component-code decoder that accepts a-priori log-likelihood ratios (LLRs), returns informative extrinsic LLRs, and remains implementable in hardware.

Importantly, the analytical advantages of algebraic component codes do not disappear in this soft-decoding regime. In many practical workflows, one first performs several iterations of soft decoding, then estimates the residual bit error probability at the component decoder input, and finally predicts the output error probability of one or several subsequent hard-decoding iterations from the known correction capability of the algebraic component code. Thus, even when the overall TPC decoder is iterative and SISO-based, the hard-decoding properties of BCH and related algebraic codes remain highly useful, especially in the final iterations. Such a methodology reduces the need for brute-force Monte Carlo simulation deep into the ultra-low-BER regime. This is especially important for optical and storage applications, where the target error rates may be so small that direct simulation is either prohibitively resource-intensive or simply infeasible in practice. For this reason, BCH and related algebraic component codes remain a particularly natural choice when one seeks both strong performance and some degree of analytical tractability.

A broad spectrum of SISO decoding algorithms for BCH codes and BCH-based TPCs has been investigated. The classical reference point is Chase--Pyndiah decoding~\cite{Pyndiah, Chase}, which combines Chase-II list generation with a simple soft-output rule and remains attractive because of its low complexity and implementation friendliness. On the other hand, optimal or near-optimal soft decoding can be obtained through trellis-based MAP algorithms, including BCJR~\cite{bcjr_1974}, symbol-wise MAP rules such as Hartmann--Rudolph decoding~\cite{hartmann_rudolph_1976}, and sectionalized- or dual-domain MAP approaches~\cite{fossorier_map_1998}. These methods provide high-quality soft information, but their complexity and memory requirements become challenging for the block lengths and throughputs relevant to practical TPC systems. Other directions explored in the literature include adaptive belief propagation for product codes~\cite{Jego2009}, Kaneko-based and other reduced-complexity list decoders, as well as heuristic post-processing and parameter-optimization techniques for Chase--Pyndiah decoding~\cite{Kim2001, Strahofer, KaiYu}.

This diversity of approaches reflects a persistent tradeoff. Methods with strong error-correction performance typically rely on trellis processing, probability-domain computations, or other operations that are difficult to map efficiently onto high-speed hardware. Conversely, methods that are easy to implement often produce degraded soft information, which limits the final coding gain of iterative decoding. In recent years, several particularly strong approaches for BCH-based TPC decoding have emerged. Among neural-network-based methods, the syndrome-based soft-output decoder of~\cite{Artemasov2025} demonstrates that learned component decoders can effectively approximate high-quality soft information for iterative algebraic-code constructions, while neural rollback~\cite{Artemasov} uses a neural model to identify and suppress unreliable extrinsic updates in Chase--Pyndiah decoding. Another state-of-the-art direction is soft-output from covered space (SOCS) decoding~\cite{Tenbrink}, which augments list-based decoding with a principled treatment of the probability mass outside the explicit candidate list, conceptually related to recent list-complement ideas in SO-GRAND~\cite{Medard}. These approaches substantially improve the quality of soft information compared with conventional Chase--Pyndiah decoding.

However, these state-of-the-art solutions still pose implementation challenges. SOCS is naturally formulated in the probability domain and requires evaluating quantities that are significantly less convenient for fixed-point, high-throughput hardware than conventional LLR-domain operations. Neural rollback, in turn, introduces an additional learned model into the decoding loop, which complicates datapath design, memory organization, quantization, verification, and certification. This motivates the search for a decoder that retains the conceptual advantages of SOCS while staying close to the max-log, list-based processing style that is more natural for hardware realizations.

In this paper, we address this problem by deriving a log-domain, max-log-oriented approximation of SOCS for iterative TPC decoding. The main contributions of this paper are as follows. First, we formulate a list-based LLR approximation for BCH component decoding that is consistent with the intuition behind SOCS but expressed directly in the log domain. Second, we introduce a normalized-offset soft-output rule parameterized by a small number of coefficients that can be optimized independently for each half-iteration. Third, we demonstrate by simulation for a TPC based on $(256,239)$ extended BCH (eBCH) component codes that the proposed decoder provides a clear gain over conventional Chase--Pyndiah decoding and offers an attractive performance--complexity tradeoff.

The remainder of the paper is organized as follows. Section~\ref{sec:system-model} describes the considered TPC transmission model. Section~\ref{sec:so-list} presents the proposed SISO decoder. Numerical results are reported in Section~\ref{sec:results}, and Section~\ref{sec:conclusion} concludes the paper.
\section{System Model}
\label{sec:system-model}

\begin{figure}[H]
    \centering
    \includegraphics{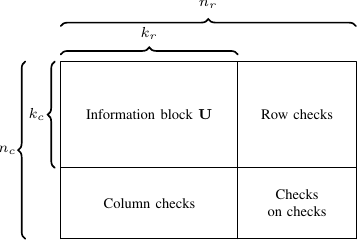}
    \caption{Turbo product code construction}
    \label{fig:tpc}
\end{figure}

Consider a communication system that employs a turbo product code based on two systematic block codes, 
$\mathcal{C}_c(n_c, k_c)$ and $\mathcal{C}_r(n_r, k_r)$. The incoming bit sequence $u$ to be encoded, of length 
$k_c \cdot k_r$, is rearranged into a rectangular matrix $\mathbf{U}$ of size $k_c \times k_r$. 
Next, for each row of this matrix, the code $\mathcal{C}_r$ performs an encoding process, 
resulting in a matrix of size $k_c \times n_r$. The same process is then applied to all columns using the 
code $\mathcal{C}_c$, yielding an $n_c \times n_r$ codeword $\mathbf{C}$. Graphical representation of encoding operation
is depicted on Fig~\ref{fig:tpc}.

The encoded binary symbols are subsequently modulated, transforming them into real-valued signals for transmission over the channel. 
In this work, we use binary phase shift keying (BPSK) modulation with the following mapping function:
\begin{equation}
    \varphi : \brc{0, 1}^n \rightarrow \brc{+1, -1}^n
\end{equation}

The result of applying modulation is a matrix $\mathbf{X}$, which is passed through the binary input additive white Gaussian noise (bi-AWGN) channel: 
\begin{equation}
    \mathbf{Y} = \mathbf{X} + \mathbf{Z},
\end{equation}
where $\mathbf{Z}$ is an additive white noise matrix, which elements are i.i.d sampled from Gaussian distribution $\mathcal{N}(0, \sigma^2)$. 
Due to the constant signal power, variance of the noise completely defines the Signal-to-Noise Ratio (SNR) of communication system as following: $E_s/N_0 = 1/2\sigma^2$.

The received noisy sequence is converted to LLRs, enabling the decoder to perform soft-decision decoding. For a given channel output assuming BPSK modulation the corresponding LLR value can be obtained as follows:
\begin{equation}
    \l_{i, j}^{\text{in}} = \ln{\frac{p(y_{i,j} \mid x_{i,j}=+1)}{p(y_{i,j} \mid x_{i,j}=-1)}} = \frac{2y_{i,j}}{\sigma^2}
\end{equation}

\section{Soft Input Soft Output Decoding}
\label{sec:so-list}

The decoding process is performed iteratively, following the reverse order of the encoding procedure. Specifically,
the decoder first processes all columns of the received matrix and then all rows.
Given a channel LLR matrix $\mathbf{L}^{\text{in}}$, the decoder iteratively approximates the a posteriori matrix $\mathbf{L}^{\text{app}}$:

\begin{equation}
    \mathbf{L}_t^{\text{app}} = \mathbf{L}^{\text{in}} + \alpha_t \mathbf{L}^{\text{ex}}_t,
    \label{eq:eq4}
\end{equation}
where $\alpha_t$ serves as a scaling factor for half-iteration $t$~\cite{Pyndiah}. By a half-iteration, we mean the update of the extrinsic information matrix $\mathbf{L}^{\text{ex}}_t$ by a component decoder using the a posteriori matrix $\mathbf{L}^{\text{app}}_{t-1}$ from the previous iteration, with $\mathbf{L}^{\text{app}}_0 = \mathbf{L}^{\text{in}}$. Since the component decoder is applied independently to each row or column, in what follows we consider the decoding of a single component codeword.

\subsection{List Decoding}

We consider a decoder that uses a list decoding algorithm to generate a set of candidate codewords.
In this work, we assume the Chase-II algorithm~\cite{Chase}.
From the received sequence of likelihoods, the decoder selects the $p$ least reliable positions and constructs a set of test patterns $\mathcal{T}$
with $2^p$ elements, representing all possible binary perturbations at those positions of the hard-decisioned received sequence $\hat{\bm{y}}$.
Let $\mathcal{I}_p$ denote the set of indices of the $p$ least reliable positions. Then,
\begin{equation}
    \mathcal{T} = \{ \hat{\bm{y}} \oplus \mathbf{e}_i, \hhquad \forall i \in [2^p]\},
\end{equation}
where $\mathbf{e}_i$ contains the bits of the binary representation of $i$ in the positions indexed by $\mathcal{I}_p$ and zeros elsewhere. Here, $[n]$ denotes the set of positive integers from $1$ to $n$.
Each generated test pattern is decoded by a bounded-distance decoder (BDD), and the set of distinct outputs forms the candidate list $\mathcal{L}$.

\subsection{Extrinsic approximation}

Once a candidate decision of the component decoder has been identified as a member of the list $\mathcal{L}$, we must evaluate its reliability in order to produce a soft output. The a posteriori LLR for a given channel output is
\begin{equation}
    l_i^{\text{app}} = \ln{\frac{P(\mathcal{C}^{(0)}_i \mid \bm{y})}{P(\mathcal{C}^{(1)}_i \mid \bm{y})}},
\end{equation}
where $\mathcal{C}_i^{(s)} \subseteq \mathcal{C}$ is the subset of codewords that have value $s$ in position $i$. Similarly, let $\mathcal{L}_i^{(s)} \subseteq \mathcal{L}$ denote the subset of list codewords that have value $s$ in position $i$. Then the probabilities can be split into two parts: those contributed by the list and those contributed by the rest of the codebook:
\begin{equation}
    l_i^{\text{app}} = \ln{\frac{P(\mathcal{L}_i^{(0)} \mid \bm{y}) + P(\mathcal{C}_i^{(0)} \setminus \mathcal{L} \mid \bm{y})}{P(\mathcal{L}_i^{(1)} \mid \bm{y}) + P(\mathcal{C}_i^{(1)} \setminus \mathcal{L} \mid \bm{y})}}
\end{equation}

Using the derivations of~\cite{Medard} with minor modifications, we introduce the following approximation for the bitwise a posteriori soft-output LLRs:
\begin{equation}
    l_i^{\text{app}} \approx \ln{\frac{P(\mathcal{L}_i^{(0)} \mid \bm{y}) + P(\mathcal{C} \setminus \mathcal{L} \mid \bm{y}_{[n] \setminus \{i\}}) \cdot P(0 \mid y_i)}{P(\mathcal{L}_i^{(1)} \mid \bm{y}) + P(\mathcal{C} \setminus \mathcal{L} \mid \bm{y}_{[n] \setminus \{i\}}) \cdot P(1 \mid y_i)}}
\end{equation}
In the original approximation, the term $P(\mathcal{C} \setminus \mathcal{L} \mid \bm{y})$ is multiplied by $P(s \mid y_i)$. In general, however, the formation of the list depends on the received sequence, so the factor $P(\mathcal{C} \setminus \mathcal{L} \mid \bm{y})$ already depends on $y_i$ through the list-generation process. Consequently, multiplying it by $P(s \mid y_i)$ would account for the information from $y_i$ twice and thus overestimate its contribution. To avoid this double counting and obtain a more transparent decomposition, we replace $P(\mathcal{C} \setminus \mathcal{L} \mid \bm{y})$ with $P(\mathcal{C} \setminus \mathcal{L} \mid \bm{y}_{[n] \setminus \{i\}})$ and condition the failure event on all received symbols except the $i$th one. After this puncturing operation, the original code reduces to a subcode whose minimum Hamming distance satisfies $d_{\text{min}} - 1 \leq d_{\text{min}}^{\prime} \leq d_{\text{min}}$.

We assume that the probability mass of each likelihood is concentrated on the most probable codeword in the corresponding set. An upper bound for these likelihoods can then be written as
\begin{equation}
    P(\mathcal{L}_i^{(s)} \mid \bm{y}) =  P(\hat{\mathbf{c}} \mid \bm{y}) + \sum_{\substack{\mathbf{c} \in \mathcal{L}_i^{(s)}, \\ \mathbf{c} \neq \hat{\mathbf{c}}}} \frac{P(\mathbf{c} \mid \bm{y})}{P(\hat{\mathbf{c}} \mid \bm{y})} \approx P(\hat{\mathbf{c}} \mid \bm{y})
\end{equation}
Therefore, the initial a posteriori LLR equation can be approximated as follows:
\begin{equation}
l_i^{\text{app}} \approx \ln{\frac{P(\hat{\mathbf{c}}^{(0)} \mid \bm{y}) + P(\widetilde{\mathbf{c}}^{(0)} \mid \bm{y}_{[n] \setminus \{i\}}) \cdot P(0 \mid y_i)}{P(\hat{\mathbf{c}}^{(1)} \mid \bm{y}) + P(\widetilde{\mathbf{c}}^{(1)} \mid \bm{y}_{[n] \setminus \{i\}}) \cdot P(1 \mid y_i)}},
\end{equation}
where $\hat{\mathbf{c}}$ and $\widetilde{\mathbf{c}}$ are the most probable codewords from the list and from the remainder of the codebook, respectively.

Assuming uniformly distributed input bits, as is common in communication systems, we factor out the ratio $\ln\frac{P(0 \mid y_i)}{P(1 \mid y_i)}$ and obtain an expression equivalent to~(\ref{eq:eq4}):
\begin{equation}
l_i^{\text{app}} \approx l_i^{in} + \ln{\frac{P(\hat{\mathbf{c}}^{(0)} \mid \bm{y}_{[n]\setminus \{i\}}) + P(\widetilde{\mathbf{c}}^{(0)} \mid \bm{y}_{[n] \setminus \{i\}})} {P(\hat{\mathbf{c}}^{(1)} \mid \bm{y}_{[n]\setminus \{i\}}) + P(\widetilde{\mathbf{c}}^{(1)} \mid \bm{y}_{[n] \setminus \{i\}})}}
\label{eq:eq11}
\end{equation}
Note that by setting the $P(\widetilde{\mathbf{c}}^{(s)} \mid \bm{y}_{[n]\setminus \{i\}})$ terms in the numerator and denominator to zero, the formula~(\ref{eq:eq11}) reduces to the Pyndiah approximation from\cite{Pyndiah}. Thus, the considered probabilities act as a regularization of the extrinsic information in the case where the original codeword is absent from the list.

Among the codewords found during the list-decoding stage, the most probable one under maximum-likelihood decoding is the one that minimizes the squared Euclidean distance to the received sequence:
\begin{equation}
\hat{\mathbf{c}} = \argmin_{\mathbf{c}\in \mathcal{L}}\lVert \bm{y} - \varphi(\mathbf{c}) \rVert_2^2
\label{eq:min}
\end{equation}

Similarly, one may argue that a codeword outside the list but inside the codebook that minimizes the same functional dominates the corresponding probability term. An exhaustive search would find such a codeword, but this procedure is effectively maximum-likelihood decoding and is too computationally demanding for real-time implementation. We therefore introduce an approximation of this probability.

To construct this approximation, we introduce a point $\widetilde{\bm{y}} \in \mathbb{R}^{n}$ as follows. Assume that the system uses maximum-likelihood decoding, which selects a codeword outside the list that minimizes the squared Euclidean distance. The point $\widetilde{\bm{y}}$ then specifies a value of this functional beyond which the corresponding codeword can be neglected because its distance from the received sequence becomes too large. In other words, although $\widetilde{\bm{y}}$ is not necessarily a codeword, it acts as an upper bound on the probability of a codeword absent from the list, for both $\widetilde{\mathbf{c}}^{(0)}$ and $\widetilde{\mathbf{c}}^{(1)}$.

For the considered list-decoding method, up to $p$ errors can be corrected through error-pattern matching, and up to $t^{\prime} = \lfloor (d_{\text{min}}^{\prime} - 1) / 2 \rfloor$ errors can be corrected by BDD. If there exists a codeword outside the list that is closest in the Euclidean metric, then it must differ from all previously considered candidates in the smallest possible number of positions. Since the Chase-II algorithm completely covers a subspace of size $2^p$ on the positions in $\mathcal{I}_p$ and all points in the Hamming ball of radius $t^{\prime}$ on the remaining positions, the target codeword must contain $t^{\prime} + 1$ errors in the remaining subspace. We therefore define the point $\widetilde{\bm{y}}$ as follows:
\begin{equation}
    \widetilde{y}_i = \begin{cases}
        -\sign{(l_i)}, \quad i \in \mathcal{I}_{p + t^{\prime} + 1} \setminus \mathcal{I}_{p} \\
        \sign{(l_i)}, \quad \text{otherwise}
        \end{cases}
        \forall i \in [n]
\end{equation}

Given $\hat{\mathbf{c}}$ and an upper bound on $\widetilde{\mathbf{c}}$ in the form of $\widetilde{\bm{y}}$, the complete a posteriori log-likelihood ratio can be expressed as

\begin{align}
l_i^{\text{ex}} 
&\approx \ln\!\left( \exp\!\left(\frac{-\lVert \bm{y} - \phi(\hat{\mathbf{c}}^{(0)}) \rVert_2^2}{2 \sigma^2}\right) 
                + \exp\!\left(\frac{-\lVert \bm{y} - \widetilde{\bm{y}}\rVert_2^2}{2 \sigma^2}\right) \right) \nonumber \\
&\quad - \ln\!\left( \exp\!\left(\frac{-\lVert \bm{y} - \phi(\hat{\mathbf{c}}^{(1)}) \rVert_2^2}{2 \sigma^2}\right) 
                + \exp\!\left(\frac{-\lVert \bm{y} - \widetilde{\bm{y}}\rVert_2^2}{2 \sigma^2}\right) \right) \nonumber \\
&= \ln\!\left(1 + \exp\!\left(\frac{-\lVert \bm{y} - \phi(\hat{\mathbf{c}}^{(0)}) \rVert_2^2 + \lVert \bm{y} - \widetilde{\bm{y}}\rVert_2^2}{2 \sigma^2}\right) \right) \nonumber \\
&\quad - \ln\!\left(1 + \exp\!\left(\frac{-\lVert \bm{y} - \phi(\hat{\mathbf{c}}^{(1)}) \rVert_2^2 + \lVert \bm{y} - \widetilde{\bm{y}}\rVert_2^2}{2 \sigma^2}\right) \right) \nonumber \\
&= \ln\!\left(1 + \exp\!\left( \frac{2\bigl(\bm{y}^T\phi(\hat{\mathbf{c}}^{(0)}) - \bm{y}^T\widetilde{\bm{y}}\bigr)}{2 \sigma^2}\right) \right) \nonumber \\
&\quad - \ln\!\left(1 + \exp\!\left( \frac{2\bigl(\bm{y}^T\phi(\hat{\mathbf{c}}^{(1)}) - \bm{y}^T\widetilde{\bm{y}}\bigr)}{2 \sigma^2}\right) \right)
\label{eq:extrinsic}
\end{align}

Due to the linear relationship between the received symbols and their corresponding log-likelihood ratios, the vector $\bm{y}$ can be replaced by twice the values of $\bm{l}^{\text{in}}$. Under the max-log approximation of~(\ref{eq:extrinsic}), this yields
\begin{equation}
l_i^{\text{ex}} \approx \max{(\Delta^{(0)}, 0)} - \max{(\Delta^{(1)}, 0)},
\label{eq:eq15}
\end{equation}
where $\Delta^{(s)}$ denotes the reliability of a codeword selected from the list for hypothesis $s$, and is defined as:
\begin{equation}
    \Delta^{(s)} = (2\bm{l}^{\text{in}})^T\phi(\hat{\mathbf{c}}^{(s)}) - (2\bm{l}^{\text{in}})^T\widetilde{\bm{y}}
    \label{eq:reliability}
\end{equation}

Intuitively, the proposed approach can be interpreted not only as introducing regularization into the extrinsic estimate when a codeword is absent from the list, but also as an approximation to decoding with covered sets~\cite{Tenbrink}. Indeed, the method examines regions of size $t^{\prime} + 1$ around the received sequence to establish a bound for the nearest codeword not contained in the list.

\begin{figure}[H]
    \centering
    \input{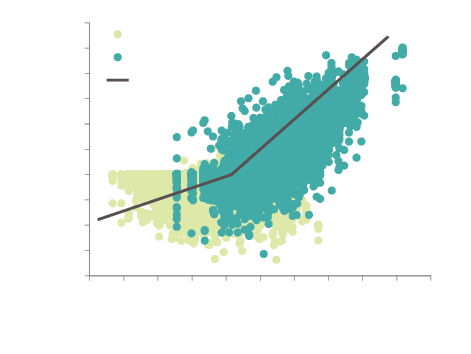}
    \caption{Correlation of extrinsic information obtained with the SOCS decoder and the proposed criterion $\Delta^{(0)}$ for all-zero codewords}
\label{fig:approximation}
\end{figure}

To verify this proposition, we devised the following experiment. We consider the all-zero codeword of the $(256,239)$ eBCH code and estimate the correlation between the extrinsic values obtained with a SOCS decoder and the reliability coefficient $\Delta^{(0)}$. Assuming the presence of an oracle, we also record whether the transmitted codeword $\mathbf{c}$ appears in the list $\mathcal{L}$. Beyond a certain reliability value, the extrinsic values exhibit approximately linear growth. Moreover, this growth occurs in the region where events are concentrated, namely when the codeword is actually present in the list. Based on these observations, the behavior of the original decoder can be approximated by a piecewise-linear function with slope coefficients $\lambda_1$ and $\lambda_2$ and offset $\mu$. This assumption constitutes a generalized version of the formula~(\ref{eq:eq15}) above and reduces to it for $\lambda_1 = 1$, $\lambda_2 = 1$, and $\mu = 0$. A numerical verification of considered experiment is shown in Fig.~\ref{fig:approximation}.

Taking the previous observation into account, when there is no alternative hypothesis, meaning that the list contains no codeword with the opposite bit value, the corresponding $\Delta$ is set to zero. This assumption is reasonable both intuitively, since in the absence of alternative hypotheses the decoder should reinforce the current hypothesis, and mathematically, because it drives the corresponding probability $P(\hat{\mathbf{c}}^{(s)} \mid \bm{y}_{[n]\setminus \{i\}})$ to zero. Motivated by this observation, we define the following function to quantify the decision reliability for the considered hypothesis:
\begin{equation}
    \psi_i^{(s)} = \begin{cases}
        0, \quad \mathcal{L}_i^{(s)}  = \varnothing \\
        \max{(\lambda_1 \cdot (\Delta^{(s)} - \mu), \lambda_2 \cdot (\Delta^{(s)} - \mu))}, \quad \text{otherwise}
    \end{cases}
\end{equation}
The corresponding extrinsic value for index $i$ is then obtained as the difference between these two quantities:
\begin{equation}
     l_{i}^{\text{ex}} \approx \psi_i^{(0)} - \psi_i^{(1)}
\end{equation}

\section{Numerical Results}
\label{sec:results}

We consider a TPC scheme based on two constituent (256, 239) eBCH codes. As the code is extended, the minimum Hamming distance of the subcode lies between 5 and 6; consequently, the error-correction capability of the subcode remains unchanged ($t = t^{\prime}$). The total number of decoding iterations is set to four. The parameters $\lambda_1$, $\lambda_2$, and $\mu$ for each half-iteration are selected via Monte Carlo simulations employing a differential evolution algorithm~\cite{Storn}. The coefficients $\alpha_t$, used for scaling the extrinsic information, are optimized using the generalized mutual information method~\cite{Strahofer}. All optimized parameters are summarized in the Table~\ref{tab:tpc-params}. In contrast to the experiment depicted in Fig.~\ref{fig:approximation}, the Monte Carlo simulations for the bit error rate (BER) curves are performed using a non-zero codeword.

\begin{table}[H]
    \centering
    \caption{Optimized parameters for numerical simulations}
    \label{tab:tpc-params}
    \begin{tabular}{c|c|c|c|c}
        \toprule
        Half iteration & $\alpha$ & $\lambda_1$ & $\lambda_2$ & $\mu$ \\
        \midrule
        1 & 0.88 & 0.47 & 0.025 & -9.22 \\
        2 & 0.86 & 0.45 & 0.027 & -10.75 \\
        3 & 0.76 & 0.43 & 0.029 & -12.28 \\
        4 & 0.74 & 0.41 & 0.031 & -13.81 \\
        5 & 0.86 & 0.39 & 0.033 & -15.35 \\
        6 & 0.82 & 0.37 & 0.035 & -16.88 \\
        7 & 0.84 & 0.36 & 0.037 & -18.41 \\
        8 & 1.00 & 0.34 & 0.039 & -19.94 \\
        \bottomrule
    \end{tabular}
\end{table}

\begin{figure}[H]
    \centering
    \includegraphics{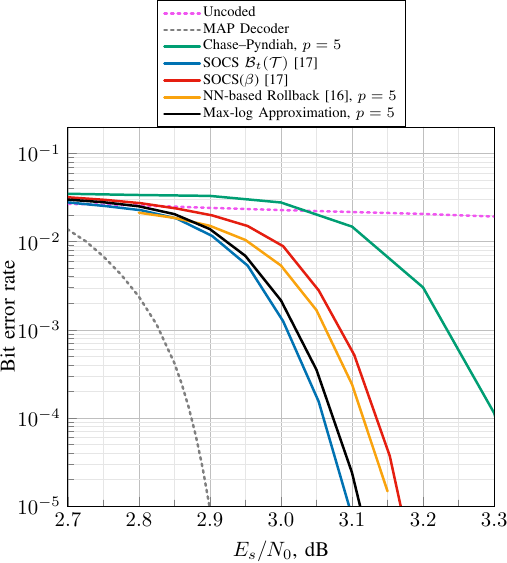}
    \caption{BER performance comparison with the state-of-the-art decoding methods}
    \label{fig:bch-results}
\end{figure}

As benchmarks, we include the MAP reference, the conventional Chase--Pyndiah decoder with $p=5$, the SOCS variants from~\cite{Tenbrink}, and the neural rollback decoder from~\cite{Artemasov}. For the SOCS decoder, we choose two configurations as comparison bounds. The SOCS($\beta$) variant serves as a upper bound, since it also relies on optimization techniques to reduce computations in the probability domain. By contrast, the SOCS decoder $\mathcal{B}_t(\mathcal{T})$ serves as an lower bound, since it fully exploits the Chase-II patterns and provides the best error-correction performance. The resulting BER curves are shown in Fig.~\ref{fig:bch-results}.

Several observations follow from Fig.~\ref{fig:bch-results}. First, the proposed decoder consistently improves upon the baseline Chase--Pyndiah decoder with the same list size. Second, the proposed approach closes a substantial part of the gap between Chase--Pyndiah and the more sophisticated SOCS-based schemes, despite relying only on log-domain operations and a small number of optimized coefficients. Third, the proposed decoder is competitive with the neural rollback approach and serves as an effective approximation of the SOCS decoder, which relies on the union of Chase-II pattern Hamming balls. At the same time, it is easier to integrate into a conventional list-decoding datapath, as it does not require a neural network inference block nor computations in the probability domain.

Overall, the results indicate that the proposed normalized offset rule offers an attractive compromise between implementation complexity and decoding performance. In particular, it inherits the low-complexity Chase-II list generation stage and replaces the probabilistic SOCS post-processing with a simple piecewise-linear mapping, which makes it a promising candidate for hardware-oriented TPC receivers.

\section{Conclusion}
\label{sec:conclusion}

This paper presented a log-domain max-log approximation of SOCS decoding for turbo product codes. Starting from the observation that the SOCS extrinsic output is strongly correlated with metric gaps between competing list hypotheses, we introduced a normalized offset decoder that maps these gaps into extrinsic LLRs through a simple piecewise-linear rule. The resulting algorithm preserves the standard Chase-based iterative decoding structure and avoids the probability-domain processing required by SOCS.

Simulation results for a TPC based on $(256,239)$ eBCH component codes showed that the proposed decoder significantly improves upon conventional Chase--Pyndiah decoding with the same list size and achieves a favorable performance--complexity tradeoff relative to recent advanced methods. Future work will focus on analytical parameter selection, hardware-oriented quantization of the proposed rule, and extensions to other list-generation algorithms and component codes.

\balance
\bibliographystyle{IEEEtran}
\bibliography{refs}

\end{document}